\documentclass[
               reprint, 
               twocolumn,
               amsmath,
               amssymb,
               showpacs,
               prl,
              superscriptaddress]{revtex4-2}

 \usepackage{ucs}
 \usepackage[utf8x]{inputenc}
 \usepackage{latexsym}
\usepackage[dvipsnames]{xcolor}
\usepackage{graphicx}
\usepackage{subcaption} 
\usepackage[rmdefault=mdput,
            OMLmathrm,
            OMLmathbf         
           ]{isomath}
\begin{document}


\author{Emilio Rodr\'iguez-Cuenca}
\email[e-mail: ]{emilioerc@gmail.com}
\affiliation{Theoretische Chemie, PCI, Universität Heidelberg, Im Neuenheimer Feld 229, D-69120 Heidelberg, Germany}

\author{Antonio Pic\'on}
\affiliation{Departamento de Qu\'{i}mica, Universidad Aut\'{o}noma de Madrid, 28049 Madrid, Spain}
\affiliation{Condensed Matter Physics Center (IFIMAC), Universidad Autónoma de Madrid, 28049, Madrid, Spain.}

\author{Solène Oberli}
\affiliation{Laboratory of Theoretical Physical Chemistry,
Institute of Chemical Sciences and Engineering, École Polytechnique Fédérale de Lausanne (EPFL), CH-1015 Lausanne, Switzerland}
\affiliation{Laboratory for Ultrafast X-ray Sciences, Institute of Chemical
Sciences and Engineering, École Polytechnique Fédérale de Lausanne (EPFL), CH-1015 Lausanne, Switzerland}
\author{Alexander I. Kuleff}
\affiliation{Theoretische Chemie, PCI, Universität Heidelberg, Im Neuenheimer Feld 229, D-69120 Heidelberg, Germany}

\author{Oriol Vendrell}
\affiliation{Theoretische Chemie, PCI, Universität Heidelberg, Im Neuenheimer Feld 229, D-69120 Heidelberg, Germany}

\title{Core-hole Coherent Spectroscopy in Molecules} 

\date{\today}

\begin{abstract}

We study the ultrafast dynamics initiated by a coherent superposition of core-excited states of nitrous oxide molecule. Using high-level \textit{ab-initio} methods, we show that the decoherence caused by the electronic decay and the nuclear dynamics is substantially slower than the induced ultrafast quantum beatings, allowing the system to undergo several oscillations before it dephases. We propose a proof-of-concept experiment using the harmonic up-conversion scheme available at X-ray free-electron laser facilities to trace the evolution of the created core-excited-state coherence through a time-resolved X-ray photoelectron spectroscopy. 

\end{abstract}

\maketitle


Direct observation of electron dynamics processes has attracted a lot of interest in the last two decades, fostered by the observation of charge-directed reactivity in peptide chains \cite{JCP99_11255(1995),JCP100_18567(1996),CPL285_25(1998)} and, from it, the emerging concept of charge migration that involves the ionization of molecules \cite{CPL285_25(1998),CPL307_205(1999)}. The possibility to initiate, monitor, and eventually control the charge-migration process has led to the paradigm of attochemistry, namely steering the chemical reactivity of a molecular system by directly controlling its electron dynamics \cite{lepine2014attosecond,nisoli2017attosecond,kuleff2018ultrafast}.

Charge migration is a pure quantum effect that results from the coherent population of electronic states, arising either due to the electron correlation \cite{CPL307_205(1999),kuleff2014ultrafast,kuleff2018ultrafast} or simply due to the broad bandwidth of the ionizing pulse \cite{woerner17charge}. The phenomenon, manifesting in ultrafast variations of the charge density throughout the molecule, has been extensively studied theoretically, see for example Refs.~\cite{lunnemann08penna,kuleff11radiation,golubev15control,Kus13pump,Mignolet14charge,Lara-Astiaso17,Lara-Astiaso18,op_dec,jia2019timing,folorunso2021molecular,Ruberti23}. First experiments have also been performed using table-top attosecond light sources \cite{Calegari2014science,kraus2015measurement,maansson2021real} and X-ray Free-Electron Lasers (XFELs) \cite{barillot21correlation}. Although typically investigated in ionized systems, charge migration initiated by electronic excitation has also been considered \cite{dutoi2010tracing,dutoi2011anExcited,yuan2019ultrafast} and recently observed experimentally \cite{matselyukh2021decoherence}, and extended to solids \cite{malakhov2023exciton}.

Although mostly studied after valence ionization or excitation, electronic coherences can be created also after core ionization \cite{kuleff2016core} or even after the core-hole relaxation in dicationic states \cite{Picon2018pra}. A first study in nitrosobenzine showed that due to strong electronic-relaxation effects, N-$1s$ ionization induces sub-femtosecond charge oscillations in the valence shell \cite{kuleff2016core}. Coherent population of core-ionized or core-excited states associated with different atoms in a molecule can also be created \cite{inhester2019jcp}. Here, we investigate such physical scenario. Analogously to the case of a quantum superposition of valence states, an x-ray pulse with a large enough bandwidth can also induce a coherent population of core-excited states. We aim at exploiting the possibility to use the recent ultrafast capabilities at XFELs facilities to resolve in time the electron dynamics triggered by an attosecond XFEL pulse. For this purpose, the N$_2$O molecule appears especially suitable. The different chemical environment of the two nitrogen atoms results in a chemical shift of about 4~eV between the core-excited (CE) states corresponding to N$_{t,c}$1s $\rightarrow$ 3$\pi^*$ transitions, where $t$ and $c$ indicate whether the excitation is from the terminal or the central nitrogen atom, respectively. A coherent population of the two excited states will thus initiate quantum beatings with a period of $T = 2\pi\hbar / \Delta E$, that is, $T\sim 1$~fs. With a decay time of 4.3~fs and 5.6~fs for N$_{t}$ and N$_{c}$, respectively \cite{Larkins1996ajp}, the core-excited transient system will be able to perform several quantum oscillations before undergoing a resonant-Auger process. How can one initiate and detect such a process?   

The development of XFEL technology has paved the way for a new era in ultrafast spectroscopy. These sources provide a unique combination of high temporal and spatial resolution with high intensity ($\sim 10^{22}$ photons/cm$^2$) \cite{Kraus2018-ml,CRYAN20221}. Recent advancements have opened the possibility to use a pump-probe harmonic up-conversion scheme in the soft X-ray regime, where the probing frequency is a harmonic of the pump frequency \cite{harmonic2,Duris2020,CRYAN20221}. This setup provides sub-femtosecond time-delay control between pulses, enabling to explore nonlinear-spectroscopy techniques for tracing ultrafast electronic dynamics. 


\begin{figure*}
  \centering
    \includegraphics[width=\textwidth]{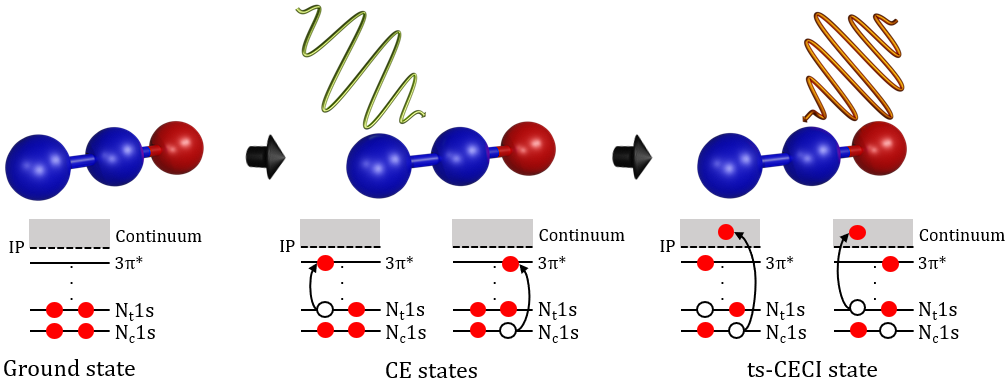} 
  \caption{Scheme of the proposed time-resolved X-ray photoelectron spectroscopy of N$_2$O}
  \label{Fig1}
\end{figure*}

We propose to use the harmonic up-conversion XFEL scheme to trace the evolution of the created core-excited-state coherence through a time-resolved X-ray photoelectron spectroscopy (TR-XPS), see figure \ref{Fig1}. An attosecond pump pulse with a carrier frequency lying between the two excitations and a bandwidth that encompasses the two CE resonances coherently excites the system and initiates electron dynamics. A delayed intense pulse of double frequency then ionizes from either of the two N-$1s$ orbitals, creating a two-hole--one-particle (2h1p) state with two core holes. The photoelectrons carry information about the electronic configuration of the system at the moment of ionization, allowing in this way to trace the evolution of the coherent CE state. 

XPS is a powerful tool for identifying the chemical state of the system constituents \cite{sieg1,sieg2}. It is element specific \cite{be_specific} and sensitive to the local chemical environment (through the chemical shift) \cite{sieg}. Although the transient absorption spectroscopy \cite{ding2021faraday,matselyukh2021decoherence} has similar selectivity, XPS might be advantageous in situations where different charge states of the studied species coexist in the sample, as the signal can be easily separated. Moreover, recent experiments have demonstrated the possibility to use XPS to track chemical bond changes in real time \cite{mayer2022nature,haddad2022nature}. However, the sensitivity of XPS to coherent electronic dynamics has not yet been proven. In this work we present a numerical experiment on N$_2$O that shows the potential of TR-XPS for tracking coherent core-hole dynamics.

The construction of the TR-XPS signal relies on an accurate computation of the electronic states involved. This was achieved using extended multi-state second-order perturbation theory (XMS-CASPT2) \cite{caspt,caspt2} on top of a Complete Active Space Self-Consistent Field (CASSCF) wave function as implemented in the OpenMolcas program package \cite{openmolcas}.
The calculations used the state-average approach
including all states within the bandwidth of the excitation pump
pulse (taken to be 4~eV), and a large Atomic Natural Orbitals (ANO-L) basis set \cite{basis} with an active space composed of 16 electrons in 12 molecular orbitals.
Further details are given in the Supplemental Material (SM). The results of these calculations are summarized in Table~\ref{tab:energy} and compared, when available, to previously reported experimental data. We see that the calculations are in a very good agreement with experiment. Importantly, the calculations show that although there are other excited states within the assumed excitation-pulse bandwidth, they have either negligibly small transition dipole moment or result from a double excitation of the system.  


\begin{table}[h!]
\caption{\label{tab:energy}Transition energies of the core-excited (CE) and the two-site core-excited-core-ionized (ts-CECI) states. Transition dipole moments ($\mu$) from the ground state to the CE states. Calculations performed at the XMS-CASPT2/ANO-L level of theory. Comparison with experimental data.}
\begin{tabular*}{\linewidth}{@{\extracolsep{\fill}}cccccc}
\hline\hline
\multicolumn{1}{l}{} & \multicolumn{3}{c}{Energy [eV]}       & \multicolumn{2}{c}{$\mu$ [a.u.]} \\ \cline{2-6}
  & CE N$_t$      & CE N$_c$       & ts-CECI & $\mu_{\mathrm{N}t}$     & $\mu_{\mathrm{N}c}$\\ \hline
Theory & 400.96    & 404.59    & 812.48    & -7.3$\times10^{-2}$ & 7.8$\times10^{-2}$  \\
Exp. \cite{exp_n2o_ce}  & 401.1  & 404.7  & - & - & - \\ \hline\hline
\end{tabular*}
\end{table}

The description of the continuum state was approached through a discretization, that is, all photoelectrons with energies $E_\text{pe}$, covering the bandwidth of the pulse (4~eV \cite{priv_comm}) with a grid step of 0.01~eV, and satisfying the energy conservation $E_\text{pe} = E_\text{probe} + E_\text{CE} - E_\text{ts-CECI}$ were taken into account.
The wave functions of these discretized states were calculated using an in-house code based on the single-center method \cite{singlecenter,singlecenter3}. This approach involves a molecular orbital expansion in spherical harmonics basis, centered on a selected single center, specifically the core-ionized N$1s$ orbital. Subsequently, the system of coupled Hartree-Fock differential equations is solved for a specific $E_\text{pe}$ value.
Given the non-centrosymmetric nature of N$_2$O, the eigenstates obtained are taken as superpositions of spherical harmonics basis functions up to $l$ and $m$ quantum numbers equal to 10, where convergence was achieved.

 \begin{figure}[ht]
  \centering
    \includegraphics[width=\linewidth]{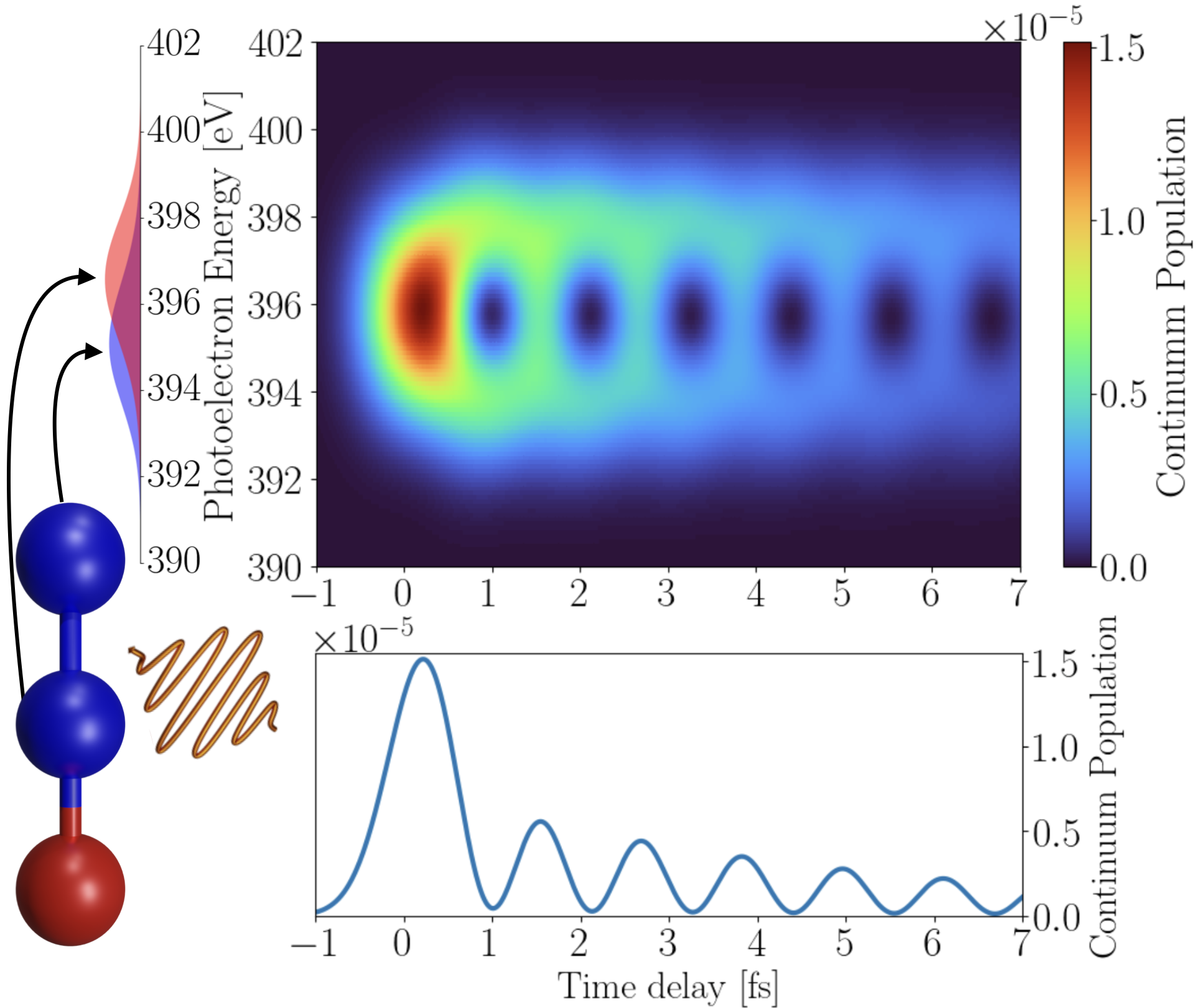} 
  \caption{Upper panel: Time-resolved XPS computed for a pulse intensity of 10$^{16}$~W/cm$^2$ at equilibrium geometry. Lower panel: Slice through the central energy of the interference.}
  \label{fig:xps16}
\end{figure}

Finally, the TR-XPS signal was obtained by evaluating the ionization probability of the CE states as a function of the time delay between the pump and probe pulses (for detailed procedure see SM). The parameters of the XFEL assumed are a pulse duration of 0.7~fs FWHM ($\sigma = 0.42$), an intensity of 10$^{16}$W/cm$^2$, and central frequencies of 402.8~eV and 805.6~eV for the pump and probe pulses, respectively. We note that these parameters are currently within reach at XFEL facilities. 

The resulting TR-XPS trace is shown in the upper panel of Fig.~\ref{fig:xps16}. We see that the trace enables to distinguish the photoelectrons emitted from the core-excited N$_t$ and N$_c$ state, as they are located at about 394 and 398~eV, respectively. This is indicated in the leftmost part of Fig.~\ref{fig:xps16}, where the two distinct photoelectron probability distributions are depicted, each corresponding to one of the two nitrogen sites at a time delay $= 0$ (no coupling terms are present). In the energy region in-between, therefore, where one cannot determine the source of photoelectrons (the probability distributions overlap), we see the quantum beats resulting from the interference between the two continuum wave functions. This interference is of course a result of the coherent population of the CE states and encodes the coherent core-electron dynamics. A slice through this region of the TR-XPS, at 396~eV, is shown in the lower panel of Fig.~\ref{fig:xps16}. We clearly see that the modulations of the signal, reflecting the alternating constructive and destructive interference, appear with a period of about 1.14~fs, as expected from the energy difference between the two CE states. The overall decrease of the signal is due to the Auger decay of the CE states. As we mentioned above, the decay is slow enough to allow the observation of the CE-state coherence.


The calculations presented above were performed at fixed nuclear geometry. It is well-known, however, that the nuclear dynamics may destroy the electronic coherence rather fast. Sometimes, it might take only 2-3~fs \cite{li_13_038302,coh2,op_dec}, which strongly limits the possibility to experimentally observe such electronic coherences. It is, therefore, necessary to examine the influence of the nuclear dynamics on the evolution of the initially prepared CE-state electronic wave packet. For this purpose, using the same level of theory XMS-CASPT2 for the electronic structure, we construct the potential-energy surfaces (PES) of the two CE states taking into account the N--N and N--O internuclear distances, thus accounting for the two fast vibrational modes of the molecule (symmetric and asymmetric stretch).
With a harmonic vibrational frequency of 596.31~cm$^{-1}$ \cite{zuniga03}, corresponding to a vibrational period of about 56~fs, and an initially linear structure,
the bending mode is too slow to play a significant role in the first few femtoseconds. However, for the sake of completeness, we incorporate it into the model, although
the stretching and bending contributions are considered separately (see below).

The nuclear dynamics in the two CE states is simulated by propagating the initial nuclear wave packet on the 2D (stretching) and 1D (bending) PESs of the two CE states. 
We assume the Condon approximation in coherent population of the two CE states
by the broad-band excitation. Thus, the initial wave packet promoted to each CE state is taken to be the vibrational ground state of the molecule.
The weights of the initial wave packets reflect the corresponding transition dipole moments obtained for the equilibrium geometry and given
in Table~\ref{tab:energy}.
The wave packets were propagated using the multiconfigurational time-dependent Hartree (MCTDH) method \cite{mctdh,mctdh1}. It is important to note that the non-adiabatic couplings between the CE states were found to be negligible in the region of interest and thus diabatization of the PES was not necessary. 


 \begin{figure}[ht]
  \centering
  \includegraphics[width=\linewidth]{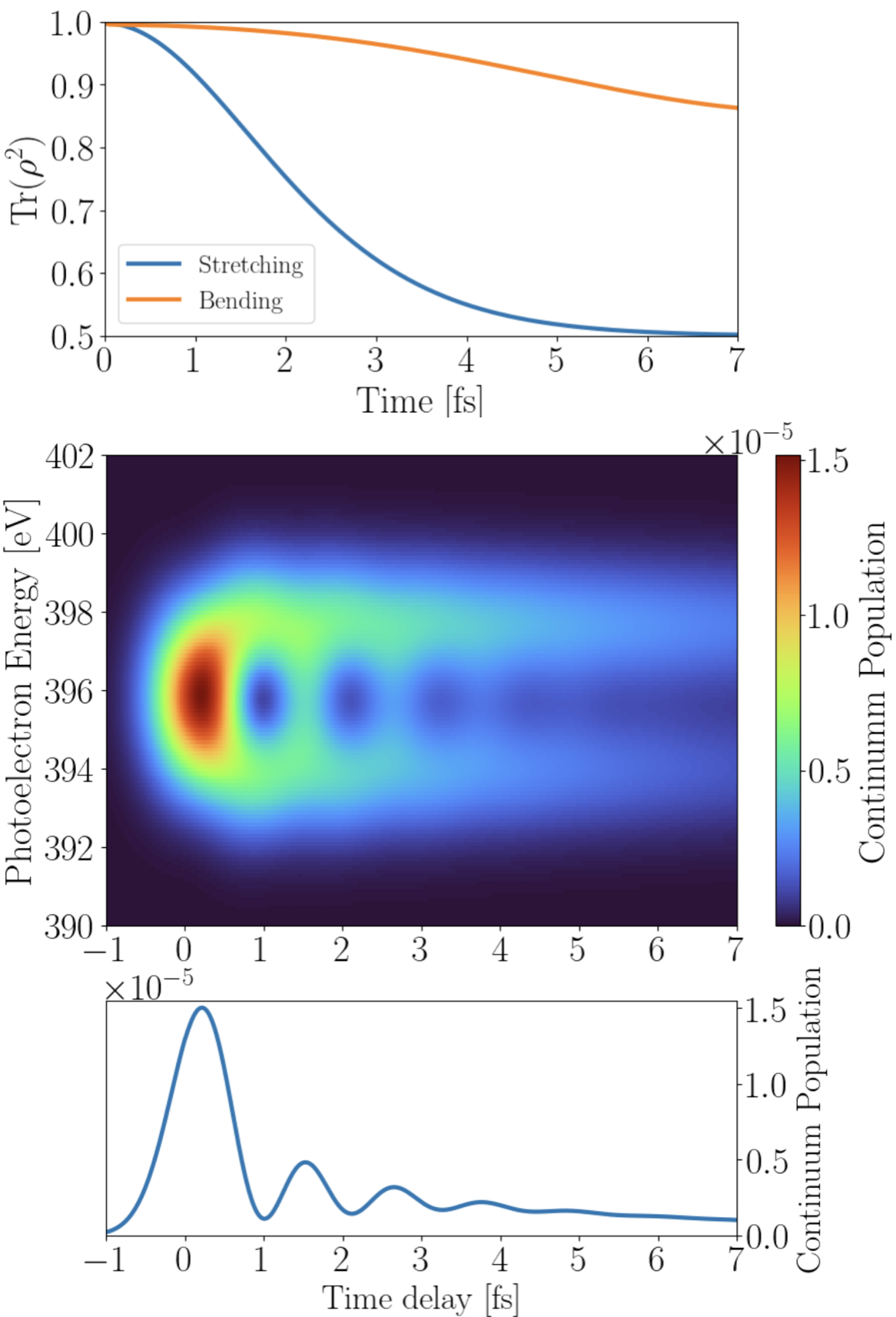}
  \caption{Upper panel: Time-dependent electronic purity of the superposition of core-excited states for the stretching and bending motions. Middle panel: Time-resolved XPS for a pulse intensity of 10$^{16}$~W/cm$^2$ including full-dimensional nuclear motion contributions. Lower panel: Slice through the central energy of the interference including full-dimensional nuclear motion contributions.}
  \label{fig:decoherence}
\end{figure}


Knowing how the nuclear wave packets propagate on the two coherently populated electronic potentials allows us to calculate the evolution of the electronic purity of the system in time. The electronic purity can be quantified using the trace of the square of the reduced density matrix, denoted as $\mathrm{Tr}(\rho^2)$. In the case of a two-state system, such as the one under investigation in this study, $\mathrm{Tr}(\rho^2) = \rho_{11}^2 + \rho_{22}^2 +2|\rho_{12}|^2$, see SM for more details. The first two terms correspond to the incoherent contributions, while the last term represents the coherence of the system. Consequently, the value of $\mathrm{Tr}(\rho^2)$ becomes 0.5 when coherence is completely lost, indicating that the system has undergone full decoherence, see  Refs.~\cite{coh1,coh2}.
When this occurs, the coherent contribution to the TR-XPS signal
disappears (see SM for details). The electronic purity is depicted in the upper panel of Fig.~\ref{fig:decoherence}. We see that the electronic coherence is kept for about 7~fs for the stretching motion, while the bending mode affects only slightly the coherence during the first few femtoseconds. We see that although the nuclear dynamics noticeably affects the created electronic coherences (compare the middle panels of Figs.~\ref{fig:xps16} and \ref{fig:decoherence}), the system still has enough time to undergo several oscillations
before the coherent oscillatory contributions disappear, as seen in the
lower panel of Fig.~\ref{fig:decoherence}. Interestingly, we could use this same methodology for more complex systems and perform a preliminary investigation of which are the nuclear degrees of freedom that would contribute more to the dephasing of the electronic beatings via the purity analysis.

Before concluding we would like to touch upon another issue that might affect the contrast of the signal. Due to the stochastic character of the pulse-generating process, the XFEL pulses exhibit shot-to-shot variations in shape, frequency, and intensity that might in principle blur the XPS trace. The influence of these variations on the charge-migration dynamics following coherent population of cationic states was thoroughly studied recently by Grell \textit{et al.}~\cite{grell2023effect}. Performing extensive computations and averaging over 100 different pulses, it was shown that the effect is negligible and the results differ only slightly from those obtained with an ideal Gaussian pulse, as used in the present calculations. Moreover, the shot-to-shot variations will change the population of core-excited states and the ionization yield intensities, but in a possible experiment this can be overcome by measuring the shot-to-shot spectra of the pulses and doing a post-selection. Hence, implementing our scheme should be feasible within the current capabilities to produce attosecond pulses at XFELs.


In summary, we demonstrate the possibility to create and measure electronic
coherences in a very short timescale of an evolving quantum system created by core-excited states. Specifically, given N$_2$O's
not extremely fast Auger decay and
relatively long vibronic coherence time,
we anticipate observing multiple oscillations using our proposed nonlinear scheme, making it a potential candidate for a proof-of-concept gas-phase experiment at an XFEL facility. Our study on N$_2$O can be further extended to other molecules, for example to organic molecules with phenyl groups, ideal candidates to create core-excited states at the carbon sites, which the sensitivity of the chemical shifts and of the double-core-hole states to the distance between their holes may be essential to discern different signals.

The main prerequisite for observing a clear signal is that the created electronic coherences have to manifest in quantum beatings with a period that is shorter than the decoherence time, which is determined by the Auger decay lifetime and the dephasing induced by the nuclear motion. The lifetime of core-excited nitrogen will be on the order of several femtoseconds in every nitrogen-containing molecule. This is also the usual timescale of core-excited carbon and oxygen. This suggests that the decay time would not be the limiting factor for a large number of bio-relevant molecules, provided the energy gap between the coherently populated states is at least about of 1~eV. The timescale of the dephasing induced by the nuclear motion is very much system- and state-dependent, and is largely unexplored. If the timescale for the core-excited states is not very different from what we know from the valence regime (from several to few tens of femtoseconds), the TR-XPS should be applicable in most of the cases. The proposed nonlinear scheme will thus provide information both on the electron relaxation and nuclear dynamics via the study of coherences, which is a fundamental and necessary step towards four-wave mixing schemes \cite{Tanaka2002,Bencivenga2015,Rouxel2021}.



\vspace{0.2cm}

We acknowledge fruitful discussions with Gilbert Grell. E. Rodr\'iguez-Cuenca thanks International Max Plank Research School for Quantum Dynamics in Physics, Chemistry and Biology (IMPRS-QD) for financial support and computing resources provided by the state of Baden-Württemberg through bwHPC. A.P. acknowledges Comunidad de Madrid through TALENTO grant refs. 2017-T1/IND-5432 and 2021-5A/IND-20959, and the Spanish Ministry of Science, Innovation and Universities \& the State Research Agency through grants refs. PID2021-126560NB-I00 and CNS2022-135803 (MCIU/AEI/FEDER, UE), and the "Mar\'ia de Maeztu" Programme for Units of Excellence in R\&D (CEX2023-001316-M), and FASLIGHT network (RED2022-134391-T), and computer resources and assistance provided by Centro de Computaci\'on Cient\'ifica de la Universidad Aut\'onoma de Madrid, Universidad de M\'alaga (FI-2022-3-0022), and Barcelona Supercomputing Center (FI-2021-3-0019,FI-2023-1-0035), and COST Action NEXT, CA22148  (European Cooperation in Science and Technology). O. Vendrell thanks the German Research Foundation for financial support through project 493826649.

\bibliography{n2o}
 
\end{document}